\documentclass[12pt,reqno,twoside]{amsart}
\usepackage{amsmath,amssymb,amsfonts,amsthm,indentfirst,setspace}
\usepackage{xcolor}\usepackage{ulem}

\newcommand{\be}{\begin{equation}}
\newcommand{\ee}{\end{equation}}
\newcommand{\bea}{\begin{eqnarray}}
\newcommand{\eea}{\end{eqnarray}}
\newcommand{\eeas}{\end{eqnarray*}}
\newcommand{\beas}{\begin{eqnarray*}}

\begin{document}

\title[Geodesic deviation equation in $f(Q)$-gravity]{Geodesic deviation equation in $f(Q)$-gravity}
\author{Jing-Theng Beh \and Tee-How Loo \and Avik De}
\address{J. T. Beh\\
Institute of Mathematical Sciences\\
Universiti Malaya\\
50603 Kuala Lumpur\\
Malaysia}
\email{bjtheng@hotmail.com}

\address{T. H. Loo\\
Institute of Mathematical Sciences\\
Universiti Malaya\\
50603 Kuala Lumpur\\
Malaysia}
\email{looth@um.edu.my}

\address{A. De\\
Department of Mathematical and Actuarial Sciences\\
Universiti Tunku Abdul Rahman\\
Jalan Sungai Long\\
43000 Cheras\\
Malaysia}
\email{de.math@gmail.com}

\footnotetext{ The authors are supported by the grant FRGS/1/2019/STG06/UM/02/6 }

\begin{abstract}
In the present paper we study the Geodesic Deviation Equation (GDE) in the modified $f(Q)$-gravity theories. The formulation of GDE in General Relativity in the case of the homogeneous and isotropic Friedman-Lema\^{i}tre-Robertson-Walker (FLRW) spacetime is briefly discussed and then extended  in modified $f(Q)$-gravity using its covariant counterpart. The generalised Mattig relation is obtained. Finally, an equivalent expression to the Dyer-Roeder equation in General Relativity in the case of $f(Q)$-gravity is presented.

\end{abstract}
\date{}
\maketitle

\section{{INTRODUCTION}}
Einstein's General Relativity (GR) is one of the most successful theories in Physics. This set of simple looking but complicated enough field equations 
\begin{equation}
R_{\mu\nu}-\frac{R}{2}g_{\mu\nu}=\kappa T_{\mu\nu}.\label{efe}
\end{equation}
has provided a remarkable narrative of the cosmological observational data \cite{will}, and created new insights into the concepts of space and time. The mathematical framework of this geometrical theory of gravity is based on (pseudo-)Riemannian geometry. It describes the properties of the gravitational field by using the curvature tensor of the spacetime. One of the fundamental equations in this theory is the geodesic deviation equation (GDE) which  provides a relation between the Riemannian curvature tensor and the relative acceleration between two nearby test particles. This equation describes the relative motion of free falling particles to bend towards or away from each other, under the impact of a gravitational field \cite{wald}. However, despite this undeniable success, increasing technological ability in modern observational cosmology posed new questions to GR. It turns out that the validity of GR might only be up to the astrophysical scales not exceeding the Solar system \cite{nojiri, brax}.

To resolve the imperfection of GR, one of the approaches is to modify the matter sector of the field equations (\ref{efe}) by adding some additional ‘dark’ components to the energy
budget of the universe, and the other one is to modify the gravitational sector. The most common modifications in the latter direction are achieved by generalizing the Einstein-Hilbert action term, precisely by replacing the Ricci scalar $R$ with an arbitrary function of some scalar produced from the Riemann or Ricci curvature tensor or some topological invariant, like the Gauss-Bonnet term or combination of both scalar-tensor terms; $f(R)$ gravity theory being the simplest and most popular one such \cite{sotiriou-faraoni, felice}. 
However, GR is formulated based on a very special and unique connection coefficient, the torsionless and metric-compatible Levi-Civita connection which can be posed as a function of the metric tensor and apparently there are other gravity theories equivalent to GR, such as the teleparallel gravity (TG) \cite{hayashi} and symmetric teleparallel gravity (STG) \cite{nester} unhindered to this special connection coefficient. Unlike GR, where the gravity is described by the curvature of spacetime, in both these theories the curvature is set to be zero, and the torsion of the connection controls the gravity in TG and the non-metricity of the connection does so in STG. As theories equivalent to GR, these two also inherit the same `dark' problem and so following the same idea to consider the dark contents of the universe as the contribution of the spacetime geometry as in $f(R)$-theory, modification of these two theories was also due. In this way, the role of Ricci scalar $R$ in GR is replaced by a scalar $T$ formed from the torsion tensor $T^\alpha_{\:\:\beta\gamma}$ in TG \cite{iorio} and by scalar $Q$ formed from the non-metricity tensor $Q_{\alpha\beta\gamma}$ in STG \cite{jimenez2} and thus modified $f(T)$ and $f(Q)$ theories were born. Both these theories have some drawback coined as ``the good and bad tetrad problem'' in TG \cite{tamanini} and ``the good and bad coordinates problem'' in STG \cite{zhao}. Like in TG, where the consistency of the theory depends on the choice of tetrad, the $f(Q)$-theory, which is the main focus of the present article, also relies on the choice of coordinates. In fact, under the constraint of vanishing curvature and torsion, we can always choose the so-called ‘coincident gauge’ in which the affine connection vanishes and take metric as the only fundamental variable, however, the theory will no longer be diffeomorphism invariant and might be inconsistent in some coordinates system. To avoid this issue, $f(Q)$ theory can be formulated in a covariant way \cite{zhao}.

As a natural extension to GR, the GDE was formulated in $f(R)$-gravity \cite{guarnizo, guarnizo-erratum}. 
Although the notion of geodesic deviation in TG is slightly different than GR in the sense that in GR, the motion of particle is described by the curvature of spacetime, so the GDE serves as the force equation, and on the other hand, in TG, the torsion appeared as a real force, namely, the tidal force; the teleparallel depiction of the gravitational interaction is totally equivalent to that of GR \cite{aldrovandi}. Thus, it is completely natural to put the force equation in TG into the form of geodesic equation in GR. In this approach, the corresponding GDE in TG can be obtained \cite{darabi}. This motivated us to investigate the covariant formulation of STG and find the GDE in the $f(Q)$-theory.

The outline of this paper is as follows: After introduction, in section 2 we reformulate the covariant version of field equations of $f(Q)$-theory. In section 3, we recapitulate the GDE in GR briefly before we plunge into the same in $f(Q)$-theory in section 4. Next we consider the ansatz of RW metric in section 5 and then further discuss the case of fundamental observer and null vector fields in this setting in section 6 and section 7, respectively. We finish with a discussion of Dyer-Roeder like equation in section 8 and a concluding section at the end. 


\section{{FIELD EQUATIONS}}
We begin with constructing a non-metric affine connection in symmetric teleparallelism, that is, $\nabla_\lambda g_{\mu\nu} \neq 0$ and we define the non-metricity tensor 
\begin{equation} \label{Q tensor}
Q_{\lambda\mu\nu} := \nabla_\lambda g_{\mu\nu} \,,
\end{equation}
so $Q_{\lambda\mu\nu} = Q_{\lambda(\mu\nu)}$. 
The associated affine connection can be expressed as
\begin{equation} \label{connc}
\Gamma^\lambda{}_{\mu\nu} := \mathring{\Gamma}^\lambda{}_{\mu\nu}+L^\lambda{}_{\mu\nu}
\end{equation}
where $\mathring{\Gamma}^\lambda{}_{\mu\nu}$ is the Levi-Civita connection from the metric 
\begin{equation*} 
\mathring{\Gamma}^\lambda{}_{\mu\nu} = \frac{1}{2} g^{\lambda\rho}(\partial_{\mu}g_{\rho\nu}+ \partial_{\nu}g_{\mu\rho} - \partial_{\rho}g_{\mu\nu})
\end{equation*}
and $L^\lambda{}_{\mu\nu}$ is called the disformation tensor.
It follows that,
\begin{equation} \label{L}
L^\lambda{}_{\mu\nu} = \frac{1}{2} (Q^\lambda{}_{\mu\nu} - Q_\mu{}^\lambda{}_\nu - Q_\nu{}^\lambda{}_\mu) \,.
\end{equation}
From the definition of non-metricity tensor, we construct two different types of non-metricity vectors,
\begin{equation*}
 Q_\mu := g^{\nu\lambda}Q_{\mu\nu\lambda} = Q_\mu{}^\nu{}_\nu \,, \qquad \tilde{Q}_\mu := g^{\nu\lambda}Q_{\nu\mu\lambda} = Q_{\nu\mu}{}^\nu \,.
\end{equation*}
Then, from (\ref{L}), we have
\begin{equation*}
L^\alpha{}_{\alpha\mu} = -\frac{1}{2}Q_{\mu} \,, \qquad L_{\mu\alpha}{}^\alpha = \frac{1}{2}Q_{\mu} - \tilde{Q}_{\mu} \,.
\end{equation*}
Moreover, we define the trace of the non-metricity tensor
\begin{equation} \label{scalarQ}
Q := g^{\mu\nu}(L^\alpha{}_{\beta\mu} L^\beta{}_{\nu\alpha} - L^\alpha{}_{\beta\alpha} L^\beta{}_{\mu\nu})
\end{equation}
and the superpotential tensor
\begin{equation} \label{P}
P^\lambda{}_{\mu\nu} := \frac{1}{4} \left( -2 L^\lambda{}_{\mu\nu} + Q^\lambda g_{\mu\nu} - \tilde{Q}^\lambda g_{\mu\nu} -\frac{1}{2} \delta^\lambda_\mu Q_{\nu} - \frac{1}{2} \delta^\lambda_\nu Q_{\mu} \right) \,.
\end{equation} 
Therefore, from (\ref{L}), (\ref{scalarQ}) and (\ref{P}), we have \cite{harko}
\begin{equation} \label{scalarQ2}
Q = Q_{\lambda\mu\nu}P^{\lambda\mu\nu} = -\frac{1}{2}Q_{\lambda\mu\nu}L^{\lambda\mu\nu} + \frac{1}{4}Q_\lambda Q^\lambda - \frac{1}{2}Q_\lambda \tilde{Q}^\lambda \,. 
\end{equation}

As discussed in \cite{harko}, the $f(Q)$-theory was constructed by using the constraints, $R^\rho{}_{\sigma\mu\nu} = 0$. That means there exist a special coordinate system such that the affine connection vanishes, $\Gamma^\lambda{}_{\mu\nu} = 0$. This situation is called the coincident gauge. Under this circumstance, the metric is the only dynamical variable. But as mentioned in \cite{zhao}, in any other coordinate such that the connection does not vanish, the evolution of the metric will be affected, and results a completely different theory. Therefore, by varying the action term
\begin{equation*}
S = \int \left[\frac{1}{2\kappa}f(Q) + \mathcal{L}_M \right] \sqrt{-g}\,d^4 x
\end{equation*}
with respect to thmetric, where $\kappa = 8\pi G$, we obtain the field equations \cite{harko}
\begin{equation} \label{FE1}
\frac{2}{\sqrt{-g}} \nabla_\lambda (\sqrt{-g}f'P^\lambda{}_{\mu\nu}) -\frac{1}{2}f g_{\mu\nu} + f'(P_{\nu\rho\sigma} Q_\mu{}^{\rho\sigma} -2P_{\rho\sigma\mu}Q^{\rho\sigma}{}_\nu) = \kappa T_{\mu\nu}
\end{equation}
where $f' = \partial f/\partial Q$. However, this equation is only valid in the coincident gauge coordinate.

On the other hand, we know that the curvature tensor can be written as
\begin{equation*}
R^\rho{}_{\sigma\mu\nu} = \partial_\mu \Gamma^\rho{}_{\nu\sigma} - \partial_\nu \Gamma^\rho{}_{\mu\sigma} + \Gamma^\rho{}_{\mu\lambda}\Gamma^\lambda{}_{\nu\sigma} - \Gamma^\rho{}_{\nu\lambda}\Gamma^\lambda{}_{\mu\sigma} \,.
\end{equation*}
Then, by using (\ref{connc}), we found that
\begin{equation*}
R^\rho{}_{\sigma\mu\nu} = \mathring{R}^\rho{}_{\sigma\mu\nu} + \mathring{\nabla}_\mu L^\rho{}_{\nu\sigma} - \mathring{\nabla}_\nu L^\rho{}_{\mu\sigma} + L^\rho{}_{\mu\lambda}L^\lambda{}_{\nu\sigma} - L^\rho{}_{\nu\lambda} L^\lambda{}_{\mu\sigma}
\end{equation*}
and so 
\begin{align*}
R_{\sigma\nu} &= \mathring{R}_{\sigma\nu} + \frac{1}{2} \mathring{\nabla}_\nu  Q_\sigma + \mathring{\nabla}_\rho L^\rho{}_{\nu\sigma} -\frac{1}{2} Q_\lambda L^\lambda{}_{\nu\sigma} - L^\rho{}_{\nu\lambda}L^\lambda{}_{\rho\sigma} \nonumber \\
R &= \mathring{R} + \mathring{\nabla}_\lambda Q^\lambda - \mathring{\nabla}_\lambda \tilde{Q}^\lambda -\frac{1}{4}Q_\lambda Q^\lambda +\frac{1}{2} Q_\lambda \tilde{Q}^\lambda - L_{\rho\nu\lambda}L^{\lambda\rho\nu} \,.
\end{align*}
Thus, from the teleparallelism, $R^\rho{}_{\sigma\mu\nu} = 0$, we obtain
\begin{equation*}
\mathring{R}_{\mu\nu} - \frac{1}{2}\mathring{R} = 2\nabla_\lambda P^{\lambda}_{\mu\nu} - \frac{1}{2}Q g_{\mu\nu} + (P_{\rho\mu\nu}Q^{\rho\sigma}{}_\sigma + P_{\nu\rho\sigma} Q_\mu{}^{\rho\sigma} -2P_{\rho\sigma\mu} Q^{\rho\sigma}{}_\nu) \,.
\end{equation*}
It follows that the field equations in (\ref{FE1}) can be rewritten as \cite{zhao}
\begin{equation} \label{FE2}
f' \mathring{G}_{\mu\nu} + \frac{1}{2}g_{\mu\nu}(f'Q - f) + 2f'' \nabla_\lambda Q P^\lambda{}_{\mu\nu} = \kappa T_{\mu\nu}
\end{equation}
where $$\mathring{G}_{\mu\nu} = \mathring{R}_{\mu\nu} - \frac{1}{2} g_{\mu\nu} \mathring{R}$$ and $T_{\mu\nu}$ is the energy-momentum tensor. This equation is in a form similar to the field equations in $f(R)$-gravity and we can proceed to find the GDE.

\section{{GEODESIC DEVIATION EQUATION IN GR}}

In this section, we depict several concepts about the geodesic deviation equation (GDE) in GR. Consider a congruence of geodesics of arbitrary casual character described by $x^\alpha (\nu,s)$, where $\nu$ is an affine parameter of geodesics, and $s$ is an indexed family of geodesics. That is, for each fixed $s$, the curve described by $\gamma_s(\nu)$ is a geodesic. Let $V^\alpha$ denote the normalized tangent vector field of the congruence, then $V^\alpha = \frac{dx^\alpha}{d\nu}$ and $V_\alpha V^\alpha = \epsilon$, where $\epsilon = +1, 0, -1$ if the geodesics are spacelike, null, or timelike respectively. Define $\eta^\alpha = \frac{dx^\alpha}{ds}$ as the deviation vector for the congruence.

Since $V^\alpha$ and $\eta^\alpha$ commutes, that is, $\mathcal{L}_V \eta^\alpha = \mathcal{L}_\eta V^\alpha$ (or $[V, \eta]^\alpha = 0$), so $\nabla_V \nabla_V \eta^\alpha = \nabla_V \nabla_\eta V^\alpha$. Then, by using the Ricci identity, $\nabla_X \nabla_Y Z^\alpha - \nabla_Y \nabla_X Z^\alpha - \nabla_{[X,Y]} Z^\alpha = \mathring{R}^\alpha{}_{\beta\gamma\delta}Z^\beta X^\gamma Y^\delta$, we obtain the GDE \cite{wald}
\begin{equation} \label{GDE in GR} 
\frac{D^2 \eta^\alpha}{D\nu^2} = -\mathring{R}^\alpha{}_{\beta\gamma\delta}V^\beta \eta^\gamma V^\delta 
\end{equation}
where $\frac{D}{D\nu}$ is the covariant derivative along the geodesic.

To further simplify, we assume the energy-momentum tensor in the form of a perfect fluid
\begin{equation} \label{T}
T_{\mu\nu} = (\rho+p)u_\mu u_\nu + pg_{\mu\nu}
\end{equation} 
where $\rho$ is the energy density and $p$ is the pressure. It follows that the trace of the energy-momentum tensor is 
\begin{equation} \label{trace T}
T = 3p - \rho \,.
\end{equation}
Recall that we have the Einstein field equations in GR (with cosmological constant) 
\begin{equation*}
\mathring{R}_{\mu\nu} -\frac{1}{2}\mathring{R} g_{\mu\nu} + \Lambda g_{\mu\nu} = \kappa T_{\mu\nu} \,.
\end{equation*}
Then, by using (\ref{T}) and (\ref{trace T}), the Ricci scalar and Ricci tensor can be expressed as
\begin{equation*}
\mathring{R} = \kappa (\rho - 3p) +4\Lambda
\end{equation*}
\begin{equation*}
\mathring{R}_{\mu\nu} = \kappa (\rho +p)u_\mu u_\nu +\frac{1}{2}[\kappa(\rho-p)+2\Lambda]g_{\mu\nu} \,.
\end{equation*}
Moreover, the Riemannian curvature tensor can also be expressed as \cite{wald}
\begin{align} \label{Riem}
\mathring{R}_{\alpha\beta\gamma\delta} = C_{\alpha\beta\gamma\delta} &+ \frac{1}{2} ( g_{\alpha\gamma}\mathring{R}_{\delta\beta}- g_{\alpha\delta}\mathring{R}_{\gamma\beta}+ g_{\beta\delta}\mathring{R}_{\gamma\alpha}- g_{\beta\gamma}\mathring{R}_{\delta\alpha} )  \nonumber \\
&- \frac{\mathring{R}}{6} \left( g_{\alpha\gamma}g_{\delta\beta} - g_{\alpha\delta}g_{\gamma\beta} \right)
\end{align}
where $C_{\alpha\beta\gamma\delta}$ is the Weyl tensor. If we consider $C_{\alpha\beta\gamma\delta} = 0$, together with $\epsilon = V_\alpha V^\alpha, E= -V_{\alpha}u^{\alpha}$, and $\eta_{\alpha}V^{\alpha} = \eta_{\alpha}u^{\alpha} = 0$, then the right hand side of GDE in (\ref{GDE in GR}) can be simplified as
\begin{equation*}
\mathring{R}^\alpha{}_{\beta\gamma\delta}V^\beta \eta^\gamma V^\delta = \bigg[ \frac{1}{3}(\kappa \rho + \Lambda)\epsilon + \frac{1}{2} \kappa (\rho +p)E^2 \bigg] \eta^\alpha \,.
\end{equation*}
This is the well-known Pirani equation \cite{ellis}.

\section{{GEODESIC DEVIATION EQUATION IN $f(Q)$-GRAVITY}}

In this section, we formulate the GDE in $f(Q)$-gravity. By contracting the field equations in (\ref{FE2}) with $g_{\mu\nu}$, we obtain the Ricci scalar
\begin{equation*}
\mathring{R} = \frac{1}{f'}(2f'Q -2f + 2f''P^{\lambda\rho}{}_\rho \nabla_\lambda Q - \kappa T) \,.
\end{equation*}
Then, substituting this Ricci scalar into (\ref{FE2}), we have the Ricci tensor
\begin{align*}
\mathring{R}_{\mu\nu} = \frac{1}{f'} \bigg[&\frac{1}{2} g_{\mu\nu} (f'Q-f+2f''P^{\lambda\rho}{}_\rho \nabla_\lambda Q - \kappa T) \nonumber \\
&-2f'' P^\lambda{}_{\mu\nu} \nabla_\lambda Q + \kappa T_{\mu\nu} \bigg] \,.
\end{align*}
Hence, by using (\ref{Riem}) and considering $C_{\alpha\beta\gamma\delta} = 0$, we found that
\begin{align*}  
\mathring{R}_{\alpha\beta\gamma\delta} =
&\frac{1}{2f'}  \bigg[   \kappa(g_{\alpha\gamma}T_{\delta\beta} - g_{\alpha\delta} T_{\gamma\beta} + g_{\beta\delta} T_{\gamma\alpha}- g_{\beta\gamma}T_{\delta\alpha}) \nonumber \\
&+ \left(\frac{f'Q}{3} - \frac{f}{3} - \frac{2\kappa T}{3} + \frac{4}{3}f'' P^{\lambda\rho}{}_\rho \nabla_\lambda Q \right)(g_{\alpha\gamma}g_{\delta\beta}-g_{\alpha\delta} g_{\gamma\beta} ) \nonumber \\
&+ (g_{\alpha\gamma}\mathcal{D}_{\delta\beta} - g_{\alpha\delta}\mathcal{D}_{\gamma\beta} + g_{\beta\delta}\mathcal{D}_{\gamma\alpha} - g_{\beta\gamma}\mathcal{D}_{\delta\alpha})f' \bigg]
\end{align*}
where
\begin{equation} \label{operator D}
\mathcal{D}_{\mu\nu} := -2P^\lambda{}_{\mu\nu} \nabla_\lambda Q \,\partial_Q \,.
\end{equation}
Assume the perfect fluid form of the energy-momentum tensor stated in (\ref{T}) and (\ref{trace T}), the above equation reduces to
\begin{align*} 
\mathring{R}_{\alpha\beta\gamma\delta} = 
&\frac{1}{2f'} \bigg[ \kappa(\rho+p) (g_{\alpha\gamma}u_\delta u_\beta - g_{\alpha\delta} u_\gamma u_\beta + g_{\beta\delta} u_\gamma u_\alpha- g_{\beta\gamma} u_\delta u_\alpha) \nonumber \\
&+ \left(\frac{f'Q}{3} - \frac{f}{3} + \frac{2\kappa\rho}{3} + \frac{4}{3}f'' P^{\lambda\rho}{}_\rho \nabla_\lambda Q \right)(g_{\alpha\gamma}g_{\delta\beta}-g_{\alpha\delta} g_{\gamma\beta}) \nonumber  \\
&+ (g_{\alpha\gamma}\mathcal{D}_{\delta\beta} - g_{\alpha\delta}\mathcal{D}_{\gamma\beta} + g_{\beta\delta}\mathcal{D}_{\gamma\alpha} - g_{\beta\gamma}\mathcal{D}_{\delta\alpha})f'  \bigg] \,.
\end{align*}
Then, contracting with $V^\beta V^\delta$ and consider $V_\alpha V^\alpha = \epsilon$, we have
\begin{align*} 
\mathring{R}_{\alpha\beta\gamma\delta} V^\beta V^\delta = 
&\frac{1}{2f'} \bigg[ \kappa(\rho+p) [g_{\alpha\gamma}(u_\beta V^\beta)^2 - 2(u_\beta V^\beta)V_{(\alpha}u_{\gamma)} +\epsilon u_\alpha u_\gamma] \nonumber \\
&+ \left(\frac{f'Q}{3} - \frac{f}{3} + \frac{2\kappa\rho}{3} + \frac{4}{3}f'' P^{\lambda\rho}{}_\rho \nabla_\lambda Q \right)(\epsilon g_{\alpha \gamma} - V_\alpha V_\gamma) \nonumber  \\
&+ [(g_{\alpha\gamma}\mathcal{D}_{\delta\beta} - g_{\alpha\delta}\mathcal{D}_{\gamma\beta} + g_{\beta\delta}\mathcal{D}_{\gamma\alpha} - g_{\beta\gamma}\mathcal{D}_{\delta\alpha})f']V^\beta V^\delta  \bigg] \,.
\end{align*}
By raising the $\alpha$ index in the Riemannian tensor and contracting with $\eta^\gamma$, we obtain
\begin{align} \label{Riem2}
\mathring{R}^\alpha{}_{\beta\gamma\delta} V^\beta \eta^\gamma V^\delta = 
&\frac{1}{2f'} \bigg[ \kappa(\rho+p) [(u_\beta V^\beta)^2 \eta^\alpha - (u_\beta V^\beta)V^\alpha (u_\gamma \eta^\gamma) \nonumber \\
&-(u_\beta V^\beta)u^\alpha (V_\gamma \eta^\gamma) +\epsilon u^\alpha (u_\gamma \eta^\gamma)] \nonumber \\
&+ \left(\frac{f'Q}{3} - \frac{f}{3} + \frac{2\kappa\rho}{3} + \frac{4}{3}f'' P^{\lambda\rho}{}_\rho \nabla_\lambda Q \right)(\epsilon \eta^\alpha -V^\alpha (V_\gamma \eta^\gamma)) \nonumber  \\
&+ [(\delta_\gamma^\alpha \mathcal{D}_{\delta\beta} - \delta^\alpha_\delta \mathcal{D}_{\gamma\beta} + g_{\beta\delta}\mathcal{D}_\gamma^\alpha - g_{\beta\gamma}\mathcal{D}_\delta^\alpha)f'] V^\beta \eta^\gamma V^\delta  \bigg] \,.
\end{align}
By using $-V_\alpha u^\alpha = E$ and $\eta_\alpha u^\alpha = \eta_\alpha V^\alpha = 0$, (\ref{Riem2}) becomes
\begin{align} \label{GDE RHS}
&\mathring{R}^\alpha{}_{\beta\gamma\delta}V^\beta \eta^\gamma V^\delta \nonumber \\
&= \frac{1}{2f'} \bigg[ \kappa(\rho+p)E^2 \nonumber + \epsilon\left(\frac{2\kappa\rho}{3} + \frac{f'Q}{3} - \frac{f}{3} + \frac{4}{3} f'' P^{\lambda\rho}{}_\rho \nabla_\lambda Q \right) \bigg] \eta^\alpha \nonumber \\
&+ \frac{1}{2f'}\bigg[ (\delta^\alpha_\gamma \mathcal{D}_{\delta\beta} - \delta^\alpha_\delta \mathcal{D}_{\gamma\beta} + g_{\beta\delta} \mathcal{D}^\alpha_\gamma - g_{\beta\gamma}\mathcal{D}^\alpha_\delta)f' V^\beta V^\delta \bigg]\eta^\gamma \,.
\end{align}

\section{{GDE with FLRW background}}
Assuming that the universe is homogeneous and isotropic, and described by the spatially flat Friedmann-Robertson-Walker (FLRW) metric, where the line element in the Cartesian coordinates is given by
\begin{equation} \label{flrw}
ds^2 = -dt^2 + a^2(t)\delta_{ij}dx^i dx^j 
\end{equation}
where $a(t)$ is the scale factor. This implies that the only non-vanishing Christoffel symbols are \cite{wald}
\begin{equation} \label{Christoffel}
\mathring{\Gamma}^l{}_{0j} = \frac{\dot{a}}{a} \delta^l_j = \mathring{\Gamma}^l{}_{j0}, \qquad \mathring{\Gamma}^0{}_{ij} = a \dot{a} \delta_{ij} 
\end{equation}
here $i,j,k,... = 1,2,3$. So the Ricci scalar can be written as
\begin{equation} \label{Ric scalar}
\mathring{R} = 6\frac{\ddot{a}}{a} + 6\left( \frac{\dot{a}}{a}\right)^2 \,.
\end{equation}
In addition, the FLRW metric is conformally flat, that is, $C_{\alpha\beta\gamma\delta} = 0$.

Alternatively, the Christoffel symbol can be expressed as
\begin{equation*}
\mathring{\Gamma}_{\lambda\mu\nu} = -\frac{\dot{a}}{a}(-u_\lambda g_{\mu\nu} + u_\mu g_{\nu\lambda} + u_\nu g_{\lambda\mu} +u_\lambda u_\mu u_\nu) \,.
\end{equation*} 
It follows that by using (\ref{flrw}), (\ref{Christoffel}) and (\ref{Ric scalar}) can be easily obtained.

Moreover, by using the spatially flat FLRW metric in (\ref{flrw}), we find that (from appendix) 
\begin{equation} \label{Q}
Q = -6H^2
\end{equation}
where $H := \frac{\dot{a}}{a}$ is the Hubble parameter. Therefore, we know that $Q$ is only time-dependent, so 
\begin{equation} \label{nabla Q}
\nabla_\lambda Q = 12 H \dot{H} u_\lambda \,.
\end{equation}
Then, by using (\ref{Q}), (\ref{nabla Q}) and the operator defined in (\ref{operator D}), we find that some of the specific terms in (\ref{GDE RHS}) can be expressed as (from appendix)
\begin{align*}
(\delta^\alpha_\gamma \mathcal{D}_{\delta\beta} - \delta^\alpha_\delta \mathcal{D}_{\gamma\beta} + g_{\beta\delta} \mathcal{D}^\alpha_\gamma - g_{\beta\gamma}\mathcal{D}^\alpha_\delta)f'V^\beta \eta^\gamma V^\delta = -24 H^2 \dot{H} f'' (2\epsilon  +E^2)\eta^\alpha 
\end{align*}
\begin{align*}
\frac{4}{3}\epsilon \eta^\alpha f'' P^{\lambda\rho}{}_\rho \nabla_\lambda Q = 48 H^2 \dot{H}f'' \epsilon \eta^\alpha \,.
\end{align*}
Therefore, (\ref{GDE RHS}) reduces to
\begin{align}  \label{GDE fQ RHS}
\mathring{R}^\alpha{}_{\beta\gamma\delta}V^\beta \eta^\gamma V^\delta = 
&\frac{1}{2f'}\bigg[ (\kappa\rho +\kappa p -24H^2 \dot{H} f'')E^2 \nonumber \\
&+ \left(\frac{2\kappa\rho}{3} +\frac{f'Q}{3} - \frac{f}{3}\right) \epsilon \bigg]\eta^\alpha
\end{align} 
which is considered to be the generalized Pirani equation. Finally, the GDE in $f(Q)$ gravity can be written as
\begin{align*}
\frac{D^2 \eta^\alpha}{D\nu^2} = 
&-\frac{1}{2f'} \bigg[ (\kappa\rho +\kappa p -24H^2 \dot{H} f'')E^2 + \left(\frac{2\kappa\rho}{3} +\frac{f'Q}{3} - \frac{f}{3}\right) \epsilon \bigg]\eta^\alpha \,.
\end{align*} 
Notice that in this GDE only the magnitude of the deviation vector $\eta^\alpha$ is changed along the geodesics, which reflects the homogeneity and isotropy of the FLRW universe. This is not the case in anistropic universes, such as Bianchi I, where the GDE also induces a change in the direction of the deviation vector, as shown in \cite{caceres}.


\section{{GDE for fundamental observers with FLRW background}}
In the case of fundamental observers, as the affine parameter, $\nu$ coincides with the proper time of the fundamental observer, $t$, we have $V^\alpha = u^\alpha$. This implies that $\epsilon = -1$ and $E = 1$.
Then, (\ref{GDE fQ RHS}) reduces to
\begin{equation} \label{GDE FO}
\mathring{R}^\alpha{}_{\beta\gamma\delta}u^\beta \eta^\gamma u^\delta = \frac{1}{f'} \left( \frac{\kappa\rho}{6} + \frac{\kappa p}{2} - \frac{f'Q}{6} + \frac{f}{6} - 12H^2 \dot{H} f'' \right) \eta^\alpha \,.
\end{equation} 
If we let $\eta^\alpha =le^\alpha$, where $e^\alpha$ is parallel propagated along $t$, then isotropy leads to 
\begin{equation*}
\frac{De^\alpha}{Dt} = 0
\end{equation*}
and so
\begin{equation*}
\frac{D^2 \eta^\alpha}{Dt^2} = \frac{d^2l}{dt^2}e^\alpha \,.
\end{equation*}
Thus, by using (\ref{GDE in GR}) and (\ref{GDE FO}), we obtain
\begin{equation*}
\frac{d^2l}{dt^2} = - \frac{1}{f'} \left( \frac{\kappa\rho}{6} + \frac{\kappa p}{2} - \frac{f'Q}{6} + \frac{f}{6} - 12H^2 \dot{H} f'' \right)l \,.
\end{equation*}
By letting $l = a(t)$, we have
\begin{equation} \label{a dotdot}
 \frac{\ddot{a}}{a} = - \frac{1}{f'} \left( \frac{\kappa\rho}{6} + \frac{\kappa p}{2} - \frac{f'Q}{6} + \frac{f}{6} - 12H^2 \dot{H} f'' \right) \,.
\end{equation}
This equation is a special case of the generalized Raychaudhuri equation. Notice that the above equation can also be obtained by the standard forms of the modified Friedmann equations in the $f(Q)$-gravity model for flat universe \cite{ayuso}
\begin{equation} \label{mod Fried}
3H^2 = \frac{1}{f'} \left[\kappa\rho + \frac{1}{2}(f'Q-f) \right] 
\end{equation}
\begin{equation*}
 2\dot{H} + 3H^2 = - \frac{1}{f'} \left[ \kappa p - \frac{1}{2}(f'Q -f) -24H^2\dot{H}f'' \right] \,.
\end{equation*}


\section{{GDE for null vector fields with FLRW background}}
In the case of past-directed null vector fields, we have $V^\alpha = k^\alpha$ with $k_\alpha k^\alpha =0 $, and so $\epsilon = 0$.
Then, (\ref{GDE fQ RHS}) becomes
\begin{equation} \label{ricci focusing}
\mathring{R}^\alpha{}_{\beta\gamma\delta}k^\beta \eta^\gamma k^\delta = \frac{1}{2f'}(\kappa\rho +\kappa p -24H^2 \dot{H}f'')E^2 \eta^\alpha \,.
\end{equation}
This equation can be explained as the Ricci focusing in $f(Q)$-gravity. If we consider $\eta^\alpha = \eta e^\alpha, \, e_\alpha e^\alpha = 1, \, e_\alpha u^\alpha = e_\alpha k^\alpha = 0 $ and $\frac{De^\alpha}{D\nu} = k^\beta \nabla_\beta e^\alpha = 0$, in which the basis is both aligned and propagated, then (\ref{GDE fQ RHS}) can be written in a new form
\begin{equation} \label{null GDE} 
\frac{d^2 \eta}{d\nu^2} = -\frac{1}{2f'} (\kappa\rho +\kappa p -24H^2\dot{H}f'')E^2 \eta \,.
\end{equation}
As in the case of GR \cite{ellis}, all past-directed null geodesics experience focusing if $\kappa(\rho +p) >0$ except the special case with the equation of state $p = -\rho$. Thus, it is clear that (\ref{null GDE}) indicates the focusing condition for the $f(Q)$-gravity, which is 
\begin{equation*}
\frac{\kappa(\rho +p)}{f'} > \frac{24H^2\dot{H}f''}{f'} \,.
\end{equation*}
After that, (\ref{null GDE}) can be expressed in term of redshift parameter, $z$. First, we write
\begin{equation*}
\frac{d}{d\nu} = \frac{dz}{d\nu} \frac{d}{dz}
\end{equation*}
which implies that
\begin{align} \label{d^2nu/dz^2}
\frac{d^2}{d\nu^2} &= \frac{dz}{d\nu} \frac{d}{dz} \left( \frac{d}{d\nu} \right) \nonumber \\
&= \left( \frac{d\nu}{dz} \right)^{-2} \left[-\left(\frac{d\nu}{dz} \right)^{-1} \frac{d^2 \nu}{dz^2}\frac{d}{dz} + \frac{d^2}{dz^2} \right] \,.
\end{align}
For the null geodesics, we have
\begin{equation*}
(1+z) = \frac{a_0}{a} = \frac{E}{E_0} \quad \longrightarrow \quad \frac{dz}{1+z} = -\frac{da}{a} 
\end{equation*}
where  $a_0 =1$ the present value of the scale factor. For the past-directed case, we set $E_0 = -1$, so 
\begin{equation*}
dz  = -(1+z) \frac{1}{a}\frac{da}{d\nu} d\nu = -(1+z) \frac{\dot{a}}{a}E d\nu = H(1+z)^2 d\nu
\end{equation*}
and so
\begin{equation*}
\frac{d\nu}{dz} = \frac{1}{ H(1+z)^2}
\end{equation*}
Consequently,
\begin{equation*} 
\frac{d^2\nu}{dz^2} = - \frac{1}{ H(1+z)^3} \left[\frac{1}{H}(1+z) \frac{dH}{dz} +2 \right]
\end{equation*}
where 
\begin{equation*}
\frac{dH}{dz} = \frac{d\nu}{dz} \frac{dt}{d\nu}\frac{dH}{dt} = - \frac{1}{H(1+z)} \frac{dH}{dt}
\end{equation*}
where we make use of $\frac{dt}{d\nu}= E = -(1+z)$. Using $\frac{\ddot{a}}{a}=\dot{H}+ H^2$ in (\ref{a dotdot}) we get
\begin{equation*}
\dot{H}= -\frac{1}{f'}\left( \frac{\kappa\rho}{6} + \frac{\kappa p}{2} - \frac{f'Q}{6} + \frac{f}{6} - 12H^2 \dot{H} f'' \right) - H^2 \,.
\end{equation*}
Hence,
\begin{align*}
\frac{d^2\nu}{dz^2} =  -\frac{3}{ H(1+z)^3} \bigg[1+ \frac{1}{3H^2 f'}\bigg( \frac{\kappa\rho}{6} + \frac{\kappa p}{2} - &\frac{f'Q}{6} + \frac{f}{6} \nonumber \\
&- 12H^2 \dot{H}f'' \bigg) \bigg] \,.
\end{align*}
Putting this equation in (\ref{d^2nu/dz^2}), we have
\begin{align*}
\frac{d^2\eta}{d\nu^2} = (H(1+z)^2)^2 \bigg[\frac{d^2\eta}{dz^2} + \frac{3}{(1+z)} &\bigg[1 + \frac{1}{3H^2 f'}\bigg( \frac{\kappa\rho}{6} +\frac{\kappa p}{2} \nonumber \\
&- \frac{f'Q}{6} + \frac{f}{6} -12H^2 \dot{H}f'' \bigg) \bigg] \frac{d\eta}{dz} \bigg] .
\end{align*}
Finally, by using (\ref{null GDE}), the null GDE can be written in the form
\begin{align} \label{null GDE in z}
\frac{d^2\eta}{dz^2} 
&+ \frac{3}{(1+z)}\left[1+ \frac{1}{3H^2f'} \left( \frac{\kappa\rho}{6} + \frac{\kappa p}{2} - \frac{f'Q}{6} + \frac{f}{6} -12H^2 \dot{H}f''\right) \right] \frac{d\eta}{dz} \nonumber \\
&+ \frac{\kappa(\rho+p) -24H^2\dot{H}f''}{2H^2(1+z)^2f'}\eta = 0 \,.
\end{align}  
Since content of the universe is the ordinary matter and the radiation, so the $\rho$ and $p$ can be be expressed as 
\begin{equation} \label{rho and p}
\kappa \rho = 3H^2_0 \Omega_{m_0}(1+z)^3 + 3H^2_0 \Omega_{r_0}(1+z)^4, \qquad \kappa p = H^2_0 \Omega_{r_0}(1+z)^4 
\end{equation}
where we use $p_m = 0$ and $p_r = \frac{1}{3}\rho_r$.
From (\ref{mod Fried}) and (\ref{rho and p}) , we could express $H^2$ as
\begin{equation} \label{H^2}
H^2 = \frac{H^2_0}{f'}[\Omega_{m_0}(1+z)^3 + \Omega_{r_0}(1+z)^4 + \Omega_{DE}]
\end{equation}
where
\begin{equation} \label{DE parameter}
\Omega_{DE} := \frac{1}{H^2_0} \left(\frac{f'Q}{6}- \frac{f}{6} \right)
\end{equation}
is the Dark Energy parameter.  
Hence, by using (\ref{rho and p}) and (\ref{H^2}), the null GDE in (\ref{null GDE in z}) can be expressed as 
\begin{equation} \label{null GDE shorform}
\frac{d^2 \eta}{dz^2} + \mathcal{P}(H,\dot{H},z) \frac{d\eta}{dz}+ \mathcal{Q}(H,\dot{H},z) \eta = 0 
\end{equation}
where 
\begin{align}  \label{func P} 
\mathcal{P}(H,\dot{H},z) = &\frac{\frac{7}{2}\Omega_{m_0}(1+z)^3 + 4\Omega_{r_0}(1+z)^4 + 2\Omega_{DE}}{(1+z)[\Omega_{m_0}(1+z)^3+ \Omega_{r_0}(1+z)^4 + \Omega_{DE}]} \nonumber \\
&- \frac{ \frac{12\dot{H}f''}{f'}[\Omega_{m_0}(1+z)^3 + \Omega_{r_0}(1+z)^4 + \Omega_{DE}]}{(1+z)[\Omega_{m_0}(1+z)^3+ \Omega_{r_0}(1+z)^4 + \Omega_{DE}]}
\end{align} 
\begin{align} \label{func Q}
\mathcal{Q}(H,\dot{H},z) = &\frac{3\Omega_{m_0}(1+z)^3 + 4\Omega_{r_0}(1+z)^4}{2(1+z)^2[\Omega_{m_0}(1+z)^3+ \Omega_{r_0}(1+z)^4 + \Omega_{DE}]} \nonumber \\
&- \frac{\frac{24\dot{H}f''}{f'}[\Omega_{m_0}(1+z)^3 + \Omega_{r_0}(1+z)^4 + \Omega_{DE}]}{2(1+z)^2[\Omega_{m_0}(1+z)^3+ \Omega_{r_0}(1+z)^4 + \Omega_{DE}]} \,.
\end{align} 
In a particular case, where $f(Q) = Q - 2\Lambda$, so $f' = 1$ and $f'' = 0$. Thus, $\Omega_{DE}$ in (\ref{DE parameter}) reduces to 
\begin{equation*}
\Omega_{DE} = \frac{1}{H^2_0} \left(\frac{Q}{6}-\frac{Q-2\Lambda}{6}\right) = \frac{\Lambda}{3H^2_0} =: \Omega_\Lambda \,.
\end{equation*}
This implies that the $H^2$ in (\ref{H^2}) becomes the same as the Friedmann equation in GR
\begin{equation*}
H^2 = H^2_0[\Omega_{m_0}(1+z)^3 + \Omega_{r_0}(1+z)^4 + \Omega_{\Lambda}]
\end{equation*}
which confirms the obtained results. Moreover, $\mathcal{P}$ (\ref{func P}) and $\mathcal{Q}$ (\ref{func Q}) turns into 
\begin{equation*} 
\mathcal{P}(z) = \frac{\frac{7}{2}\Omega_{m_0}(1+z)^3 + 4\Omega_{r_0}(1+z)^4 + 2\Omega_\Lambda}{(1+z)[\Omega_{m_0}(1+z)^3 + \Omega_{r_0}(1+z)^4 + \Omega_\Lambda]}
\end{equation*}
\begin{equation*}
\mathcal{Q}(z) = \frac{3\Omega_{m_0}(1+z) + 4\Omega_{r_0}(1+z)^2}{2[\Omega_{m_0}(1+z)^3 + \Omega_{r_0}(1+z)^4 + \Omega_\Lambda]} .
\end{equation*} 
Then, the GDE for null vector fields becomes
\begin{align*}
\frac{d^2\eta}{dz^2} 
&+ \frac{\frac{7}{2}\Omega_{m_0}(1+z)^3 + 4\Omega_{r_0}(1+z)^4 + 2\Omega_\Lambda}{(1+z)[\Omega_{m_0}(1+z)^3 + \Omega_{r_0}(1+z)^4 + \Omega_\Lambda]} \frac{d\eta}{dz} \nonumber \\
&+ \frac{3\Omega_{m_0}(1+z) + 4\Omega_{r_0}(1+z)^2}{2 [\Omega_{m_0}(1+z)^3 + \Omega_{r_0}(1+z)^4 + \Omega_\Lambda]}\eta = 0 \,.
\end{align*}
We set $\Omega_\Lambda = 0$ and $\Omega_{m_0}+ \Omega_{r_0} =1$ for the original Mattig relation, so we have 
\begin{align*}
\frac{d^2\eta}{dz^2} 
&+ \frac{\frac{7}{2}\Omega_{m_0}(1+z)^3 + 4\Omega_{r_0}(1+z)^4}{(1+z)[\Omega_{m_0}(1+z)^3 + \Omega_{r_0}(1+z)^4 ]} \frac{d\eta}{dz}  \nonumber \\
&+ \frac{3\Omega_{m_0}(1+z) + 4\Omega_{r_0}(1+z)^2}{2[\Omega_{m_0}(1+z)^3 + \Omega_{r_0}(1+z)^4]}\eta = 0 \,.
\end{align*}
This gives us a hint that the generalized Mattig relation in $f(Q)$-gravity can be obtained from (\ref{null GDE shorform}). In FLRW universe, the angular diametral distance $D_A(z)$ is given by \cite{schneider}
\begin{equation*}
D_A(z) = \sqrt{\left|\frac{dA(z)}{d\Omega}\right|}
\end{equation*}
where $dA$ is the area of the object and $d\Omega$ is the solid angle. Thus, from (\ref{null GDE shorform}), the GDE in terms of the angular diametral distance is
\begin{equation} \label{GDE ADD}
\frac{d^2 D_A}{dz^2} + \mathcal{P}(H, \dot{H}, z) \frac{d D_A}{dz} + \mathcal{Q}(H, \dot{H}, z) D_A = 0
\end{equation} 
where $\mathcal{P}$ and $\mathcal{Q}$ is given in (\ref{func P}) and (\ref{func Q}). This equation satisfies the initial conditions (for $z \geq z_0$)
\begin{align} 
D_A (z)|_{z=z_0} &= 0 \label{int cond 1} \\
\frac{d D_A}{dz} (z)|_{z=z_0} &= \frac{H_0}{H(z_0)(1+z_0)} \label{int cond 2}
\end{align}  
where $H(z_0)$ is the modified Friedmann equation (\ref{H^2}) at $z=z_0$.

\section{Dyer-Roeder like equation in $f(Q)$-gravity}
Finally we describe a relation that is a tool to investigate cosmological distances in inhomogeneous universes. The Dyer-Roeder equation is a differential equation for the diametral angular distance as a function of the redshift. The Dyer-Roeder equation in GR is given by \cite{okamura}
\begin{equation*}
(1+z)^2 \mathcal{F}(z) \frac{d^2 D_A}{dz^2} + (1+z) \mathcal{G}(z) \frac{dD_A}{dz^2} + \mathcal{H}(z) D_A = 0
\end{equation*}
where 
\begin{equation*}
\mathcal{F}(z) = H^2(z) \,,
\end{equation*}
\begin{equation*}
\mathcal{G}(z) = (1+z) H(z) \frac{dH}{dz} + 2H^2(z) \,,
\end{equation*}
and
\begin{equation*}
\mathcal{H}(z) = \frac{3\tilde{\alpha}(z)}{2} \Omega_{m0} (1+z)^3 \,,
\end{equation*}
where $\tilde{\alpha}(z)$ is the smoothness parameter, which provides the property of inhomogeneities in the energy density. The influence of the smoothness parameter $\tilde{\alpha}$ in the behavior of $D_A(z)$ is discussed in \cite{schneider, okamura}. Now, we express the Dyer-Roeder like equation in $f(Q)$-gravity. First, notice that the terms involving the derivatives of $D_A$ in (\ref{GDE ADD}) are from the transformation $\frac{d}{d\nu} \rightarrow \frac{d}{dz}$, while the terms with $D_A$ are from the Ricci focusing in (\ref{ricci focusing}). Then, define a mass-fraction $\tilde{\alpha}$ of matter in the universe, and replacing the $\rho$ in the Ricci focusing with $\tilde{\alpha}\rho$. Hence, from (\ref{null GDE shorform}), and consider the case $\Omega_{r_0} = 0$, we obtain
\begin{align} \label{DRD}
\frac{d^2 D_A^{DR}}{dz^2} &+ \frac{\frac{7}{2}\Omega_{m_0}(1+z)^3 +2 \Omega_{DE} -\frac{12\dot{H}f''}{f'}[\Omega_{m_0}(1+z)^3+\Omega_{DE}]}{(1+z)[\Omega_{m_0}(1+z)^3 +\Omega_{DE}]} \frac{d D_A^{DR}}{dz} \nonumber \\
&+ \frac{3\tilde{\alpha}(z)\Omega_{m_0}(1+z)^3 - \frac{24\dot{H}f''}{f'}[\Omega_{m_0}(1+z)^3 +\Omega_{DE}]}{2(1+z)^2[\Omega_{m_0}(1+z)^3 +\Omega_{DE}]} d D_A^{DR} = 0
\end{align}
where $D_A^{DR}$ denote the Dyer-Roeder distance in $f(Q)$-gravity. This equation also satisfies the conditions stated in (\ref{int cond 1}) and (\ref{int cond 2}). In the case of $f(Q) = Q - 2\Lambda$, (\ref{DRD}) reduces to the standard form of GR.


\section{{Conclusion}}
In the core of this paper lies the Ricci curvature tensor corresponding to the Levi-Civita connection, expressed in terms of the tensor $Q_{\mu\nu\lambda}$ with the covariant form of the field equations of $f(Q)$-gravity theory. In the FLRW universe, the GDE corresponding to these GR comparable terms of $f(Q)$-gravity is acquired for the fundamental observer and the past-directed null vector fields. The null vector field case provides an important results which is the generalisation of the Mattig relation and the differential equation for the diametral angular distance in $f(Q)$-gravity. As an extension, the Dyer-Roeder equation was considered.

\section{Appendix}
From (\ref{Q tensor}), we can get all the non-metricity tensors as follow
\begin{align*}
Q_{\lambda\mu\nu} &= \nabla_{\lambda}g_{\mu\nu} \\
Q^\lambda{}_{\mu\nu} &= g^{\lambda\rho}Q_{\rho\mu\nu} = g^{\lambda\rho} \nabla_{\rho}g_{\mu\nu} = \nabla^{\lambda}g_{\mu\nu} \\
Q_\lambda{}^\mu{}_\nu &= g^{\mu\rho}Q_{\lambda\rho\nu} = g^{\mu\rho}\nabla_\lambda g_{\rho\nu} = -g_{\rho\nu}\nabla_\lambda g^{\mu\rho} \\
Q_{\lambda\mu}{}^\nu &= g^{\nu\rho}Q_{\lambda\mu\rho} = g^{\nu\rho}\nabla_\lambda g_{\mu\rho} = -g_{\mu\rho} \nabla_\lambda g^{\nu\rho} \\
Q^{\lambda\mu}{}_{\nu} &= g^{\lambda\rho}g^{\mu\gamma} \nabla_\rho g_{\gamma\nu} = g^{\mu\gamma}\nabla^\lambda g_{\gamma\nu} = -g_{\gamma\nu}\nabla^\lambda g^{\mu\gamma} \\
Q^\lambda{}_\mu{}^\nu &= g^{\lambda\rho}g^{\nu\gamma} \nabla_\rho g_{\mu\gamma} = g^{\nu\gamma} \nabla^\lambda g_{\mu\gamma} = -g_{\mu\gamma} \nabla^\lambda g^{\nu\gamma} \\
Q_\lambda{}^{\mu\nu} &= g^{\mu\rho}g^{\nu\gamma}\nabla_\lambda g_{\rho\gamma} = -g^{\mu\rho}g_{\rho\gamma} \nabla_\lambda g^{\nu\gamma} = -\nabla_\lambda g^{\mu\nu} \\
Q^{\lambda\mu\nu} &= - \nabla^\lambda g^{\mu\nu} \,.
\end{align*}

Recall in (\ref{scalarQ2}), we have
\begin{equation*}
Q = -\frac{1}{4} Q_{\lambda\mu\nu}Q^{\lambda\mu\nu} + \frac{1}{2}Q_{\lambda\mu\nu} Q^{\mu\lambda\nu} +\frac{1}{4}Q_\lambda Q^\lambda - \frac{1}{2}Q_\lambda \tilde{Q}^\lambda \,.
\end{equation*}

By using the above results and the FLRW metric in (\ref{flrw}), we obtain
\begin{align*}
Q_{\lambda\mu\nu}Q^{\lambda\mu\nu} &= - \nabla_\lambda g_{\mu\nu} \nabla^\lambda g^{\mu\nu} = -12 H^2 \\
Q_{\lambda\mu\nu}Q^{\mu\lambda\nu} &= -\nabla_\lambda g_{\mu\nu} \nabla^\mu g^{\lambda \nu} = 0 \\
Q_\lambda Q^\lambda &= (g_{\mu\rho}\nabla_\lambda g^{\mu\rho})(g_{\nu\gamma} \nabla^\lambda g^{\nu\gamma}) = -36 H^2 \\
Q_\lambda \tilde{Q}^\lambda &= (g_{\mu\rho} \nabla_\lambda g^{\mu\rho})(\nabla_\nu g^{\lambda\nu}) = 0 \,.
\end{align*}
Hence, we have
\begin{equation*}
Q = -\frac{1}{4}(-12H^2) +\frac{1}{4}(-36H^2) = -6H^2 \,.
\end{equation*}
Next, we try to simplify the terms 
\begin{equation*}
\frac{1}{2f'}\bigg[ (\delta^\alpha_\gamma \mathcal{D}_{\delta\beta} - \delta^\alpha_\delta \mathcal{D}_{\gamma\beta} + g_{\beta\delta} \mathcal{D}^\alpha_\gamma - g_{\beta\gamma}\mathcal{D}^\alpha_\delta)f' V^\beta V^\delta \bigg]\eta^\gamma 
\end{equation*}
\begin{equation*}
\frac{4}{3} \epsilon \eta^\alpha f'' P^{\lambda\nu}{}_\nu \nabla_\lambda Q
\end{equation*}
as stated in (\ref{GDE RHS}). From the definition of the operator $\mathcal{D}_{\mu\nu}$ in (\ref{operator D}), we have
\begin{align*}
\delta^\alpha_\gamma V^\beta V^\delta \eta^\gamma \mathcal{D}_{\delta\beta} f' &= -2 \eta^\alpha V^\delta V^\beta P^\lambda{}_{\delta\beta} \nabla_\lambda Q f'' \\
-\delta^\alpha_\delta V^\beta V^\delta \eta^\gamma \mathcal{D}_{\gamma\beta} f' &= 2 V^\alpha \eta^\gamma V^\beta P^\lambda{}_{\gamma\beta} \nabla_\lambda Q f'' \\
g_{\beta\delta} V^\beta V^\delta \eta^\gamma \mathcal{D}^\alpha_\gamma f' &= -2 \epsilon \eta^\gamma P^{\lambda\alpha}{}_\gamma \nabla_\lambda Q f'' \\
-g_{\beta\gamma} V^\beta V^\delta \eta^\gamma \mathcal{D}^\alpha_\delta f' &= 0 \,.
\end{align*}
Since $Q$ is only time-dependent, so the summation of the index $\lambda$ reduces to only the 0 component. Then, we verify the terms as follow
\begin{align*}
P^0{}_{\mu\nu} &= \frac{1}{4} \left(-Q^0{}_{\mu\nu}+ Q_\mu{}^0{}_\nu +Q_\nu{}^0{}_\mu +Q^0 g_{\mu\nu} -\tilde{Q}^0 g_{\mu\nu}-\frac{1}{2}\delta^0_\mu Q_\nu -\frac{1}{2}\delta^0_\nu Q_\mu \right) \nonumber \\
&= \frac{1}{4} \left(\nabla_0 g_{\mu\nu} -6Hg_{\mu\nu}+\frac{1}{2}\delta^0_\mu g_{\alpha\beta} \nabla_\nu g^{\alpha\beta} +\frac{1}{2}\delta^0_\nu g_{\alpha\beta} \nabla_\mu g^{\alpha\beta}\right) \\
P^{0\mu}{}_\nu &= \frac{1}{4} \left( -Q^{0\mu}{}_\nu +Q^{\mu 0}{}_\nu +Q_\nu{}^{0\mu} +Q^0 \delta^\mu_\nu -\tilde{Q}^0 \delta^\mu_\nu -\frac{1}{2}g^{0\mu}Q_\nu -\frac{1}{2} \delta^0_\nu Q^\mu \right) \nonumber \\
&= \frac{1}{4} \left(-g_{\rho\nu} \nabla_0 g^{\mu\rho} -6H \delta^\mu_\nu +\frac{1}{2}g^{0\mu}g_{\alpha\beta}\nabla_0 g^{\alpha\beta} +\frac{1}{2} \delta^0_\nu g_{\rho\nu}g^{\alpha\mu} \nabla_\alpha g^{\rho\nu} \right) \\
P^{0\nu}{}_\nu &= \frac{1}{4} \left(-g_{\rho\nu} \nabla_0 g^{\rho\nu} -6H \delta^\nu_\nu +\frac{1}{2}g^{0\nu}g_{\alpha\beta}\nabla_0 g^{\alpha\beta} +\frac{1}{2} g^{0\nu} g_{\rho\nu} \nabla_\nu g^{\rho\nu} \right) \nonumber \\
&= \frac{1}{4}(2g_{\alpha\beta} \nabla_0 g^{\alpha\beta}) \nonumber \\
&= -3H \,.
\end{align*}
Notice that if $\mu \neq \nu$, then $P^0{}_{\mu\nu} = 0$. This implies that 
\begin{align*}
V^\mu V^\nu P^0{}_{\mu\nu} &= V^0V^0 P^0{}_{00} + \sum_k V^k V^k P^0{}_{kk} \\
&= \frac{1}{4} \left[ V^0V^0(\nabla_0 g_{00}- 6Hg_{00} -6H) + \sum_k V^k V^k (\nabla_0 g_{kk} -6H g_{kk}) \right] \\
&= \frac{1}{4} \left[ V^0 V^0 (0) + \sum_k V^k V^k \nabla_0 g_{kk} - (6H) \sum_k V^k V^k g_{kk} \right] \\
&= \frac{1}{4} \left[ \sum_k V^k V_k g^{kk}\nabla_0 g_{kk} - (6H) \sum_k V^k V_k  \right]  \\
&= \frac{1}{4} \left[ \sum_k V^k V_k (2H-6H) \right] \\
&= - \frac{1}{4} (\epsilon - V_0 V^0) 4H \\
&= -H(\epsilon + E^2) \\
\eta^\mu V^\nu P^0{}_{\mu\nu} &= \eta^0 V^0 P^0{}_{00} + \sum_k \eta^k V^k P^0{}_{kk} \\
&= \frac{1}{4} \sum_k \eta^k V^k (\nabla_0 g_{kk} - 6H g_{kk}) \\
&= \frac{1}{4} \sum_k \eta^k V_k  (2H-6H) \\
&= 0 \\
\eta^\nu P^{0\mu}{}_\nu &= \eta^0 P^{0\mu}{}_0 + \eta^i P^{0\mu}{}_i \\
&= \frac{1}{4} \eta^i (-g_{\rho i} \nabla_0 g^{\rho\mu} - \delta^\mu_i 6H) \\
&= \frac{1}{4} (-\eta_\rho \nabla_0 g^{\rho\mu} - \eta^\mu 6H) \\
&= \frac{1}{4} (2\eta^\mu H - \eta^\mu H) \\
&= -H \eta^\mu \,.
\end{align*}
Thus, we have
\begin{align*}
-2 \eta^\alpha V^\delta V^\beta P^\lambda{}_{\delta\beta} \nabla_\lambda Q f'' 
&= -2\eta^\alpha (-H)(\epsilon+E^2)(-12H \dot{H})f'' \\ 
&= -24 H^2 \dot{H} f'' (\epsilon + E^2) \eta^\alpha \\
2 V^\alpha \eta^\gamma V^\beta P^\lambda{}_{\gamma\beta} \nabla_\lambda Q f'' 
&= 0 \\
-2\epsilon \eta^\gamma P^{\lambda\alpha}{}_\gamma \nabla_\lambda Q f'' 
&= -2 \epsilon (-H\eta^\alpha)(-12 H \dot{H}) f'' \\
&= -24 H^2 \dot{H} f'' \epsilon \eta^\alpha \,.
\end{align*}
Therefore,
\begin{equation*}
\frac{1}{2f'}\bigg[ (\delta^\alpha_\gamma \mathcal{D}_{\delta\beta} - \delta^\alpha_\delta \mathcal{D}_{\gamma\beta} + g_{\beta\delta} \mathcal{D}^\alpha_\gamma - g_{\beta\gamma}\mathcal{D}^\alpha_\delta)f' V^\beta V^\delta \bigg]\eta^\gamma = \frac{1}{2f'}[-24 H^2 \dot{H} f'' (2\epsilon + E^2)] \eta^\alpha
\end{equation*}
and 
\begin{align*}
\frac{4}{3} \eta^\alpha f'' P^{\lambda\nu}{}_\nu \nabla_\lambda Q 
&= \frac{4}{3} \epsilon \eta^\alpha f'' (-3H) (-12H \dot{H}) \nonumber \\
&=48 H^2 \dot{H} f'' \epsilon \eta^\alpha \,.
\end{align*}

\end{document}